\documentclass[showpacs,preprintnumbers,amsmath,amssymb]{revtex4}
\makeatletter
\begin{document}
\title{A Geometrical Approach towards Entanglement}
\author{B. Basu}
 \email{banasri@isical.ac.in}
\author{P. Bandyopadhyay}
 \email{pratul@isical.ac.in}
 \affiliation{Physics and Applied Mathematics Unit\\
 Indian Statistical Institute\\
 Kolkata-700108 }

\begin{abstract}
We have studied the concurrence of two-site entanglement and have shown that it is related to the geometric phase accumulated due to a complete rotation of the entangled state. The geometric phase and hence the concurrence is evaluated for transverse Ising model and antiferromagnetic chain which is found to be in good agreement with that obtained by other methods. 

\end{abstract}
 \pacs{03.65.Ud, 03.65.Vf }
\maketitle
\section{Introduction}
 Entanglement is one of the most striking features of quantum mechanics. It is
known long ago that quantum mechanics exhibits very peculiar
correlation between two physically distant parts of the total
system. The discovery of Bell inequality(BI)
\cite{Bell} showed that BI can be violated in quantum mechanics but
have to be satisfied by all local realistic theories. The violation
of BI demonstrates the presence of entanglement.

Evidently this  characteristic feature of quantum mechanics is induced through the quantization procedure. Klauder \cite {2} has pointed out that quantization is governed by geometry. It is shown that \cite{3} quantization can be achieved in terms of a universal magnetic field acting on a free particle moving in a higher dimensional space when quantization corresponds to freezing the particle in its lowest Landau level. The most significant aspect of this quantization procedure is its explicit property of coordinate independence thus specifying the role of geometry. The equivalence of this formalism with geometric quantization \cite{4} has also been pointed out. Again, it has been shown
\cite{5,6} that this procedure has its relevance in stochastic quantization of a fermion when a spinning particle is endowed with an internal degree of freedom through a $direction$ $vector$ (vortex line) which is topologically equivalent to a magnetic flux line. So it is expected that entanglement of a two qubit system is related to this internal degree of freedom manifested through the magnetic flux line. 

In a classic paper Berry \cite{7} has shown  that a quantum particle attains a geometric phase when it moves adiabatically in a closed path in a parameter space. This geometric phase is associated with this internal degree of freedom of a spinning particle and is given by the integral over a closed path of the corresponding gauge potential. 
In view of this, a $geometrical$ $interpretation$ of $entanglement$ in terms of this $geometric$ $phase$ is naturally expected. This is the motivation of our present paper. Indeed, the Berry phase (geometric phase) in an entangled spin 1/2 system has been studied by several authors \cite{8,9,10,11,12}. A possible relation between the Berry phase and the measure of entanglement viz. concurrence has been pointed out in ref. \cite{13}. 
Here we shall try to quantify concurrence in terms of geometric phase in several spin systems.   
 

\section{Concurrence and Geometric phase }
As is well known, a fermion is realized when a scalar particle is
attached with a magnetic flux quantum. Indeed the attachment of a
magnetic flux quantum changes the spin and statistics of the
particle. The attachment of a magnetic flux quantum to a particle
may be visualized such that the particle is moving in the field of a
magnetic monopole of strength $\mu$ which gives rise to the phase
$e^{i2\pi \mu}$\cite{dp}. It is noted that $\mu$ can take half integer
or integer values and $\mu=1/2$ corresponds to one flux quantum. 
When a scalar particle moves around a monopole of strength $\mu=1/2$ the internal degree of freedom of the spinning particle with spin $1/2$ is manifested through the direction vector which corresponds to the spin axis and is related to the orientation of the magnetic flux line.  


To study the Berry phase effect on an entangled system of two
spin $1/2$ particles
we start  with an
entangled system of two spin $1/2$ particles given by the state
\begin{equation}\label{Bel}
\psi_{-}=a |\uparrow\downarrow>-b |\downarrow\uparrow>
\end{equation}
where $a$ and $b$ are complex coefficients.
Under the influence of another spin-$1/2$ particle the magnetic flux line
in the configuration of a given fermion will change its direction. This may be considered
as the magnetic field rotating with an angular velocity
$\omega_0$ around the $z$-axis under an arbitrary angle $\theta$. Indeed we may
consider the time dependent magnetic field given by
\begin{equation}\label{Bt}
    {\bf B}(t)=B {\bf n}(\theta,t)
\end{equation}
when the unit vector ${\bf n}(\theta,t)$ may be depicted as

\begin{equation}\label{nn}
    {\bf n}(\theta,t)=
    \left(
    \begin{array}{cc}
    \sin \theta & \cos(\omega_0 t) \\
    \sin \theta & \sin(\omega_0 t) \\
    \cos \theta
    \end{array}\right)
\end{equation}

The interaction in this system is described by the
Hamiltonian
\begin{equation}\label{h}
 H=\frac{k}{2} B {\bf n}.\vec{\sigma}
\end{equation}
where $\vec{\sigma}$ are Pauli matrices and $k=g \mu_B$, $\mu_B$
being the Bohr magneton and $g$ is the Lande factor.

The instantaneous eigenstates of a spin operator in direction
 ${\bf n}(\theta,t)$ expanded in the $\sigma_z$-basis are given by
\begin{equation}\label{art}
\begin{array}{ccc}
    \displaystyle{|\uparrow_n;t>}&=&\displaystyle{\cos \frac{\theta}{2} |\uparrow_z> +~ \sin
    \frac{\theta}{2} e^{i\omega_0 t}|\downarrow_z> }\\
    &&\\
    \displaystyle{|\downarrow_n;t>}&=&\displaystyle{\sin \frac{\theta}{2} |\uparrow_z> +~ \cos
    \frac{\theta}{2} e^{i\omega_0 t}|\downarrow_z>}
\end{array}
\end{equation}

For the time  evolution from $t=0$ to $t=\tau$ where
$\tau=\displaystyle{\frac{2\pi}{\omega_0}}$ each eigenstate will
pick up a geometric phase (Berry phase) apart from the dynamical
phase which is of the form \cite{8}
\begin{eqnarray}\label{upn}
\displaystyle{|\uparrow_n;t=0>\rightarrow|\uparrow_n;t=\tau>=e^{i\gamma_+
(\theta)}
~e^{i\nu_+} |\uparrow_n;t=0>} \nonumber \\
\nonumber \\
\displaystyle{|\downarrow_n;t=0>\rightarrow|\downarrow_n;t=\tau>=e^{i\gamma_-
(\theta)} ~e^{i\nu_-} |\downarrow_n;t=0>}
\end{eqnarray}
where $\gamma_\pm$ is the Berry phase which is half of the solid
angle $\frac{1}{2}\Omega$ swept out by the magnetic flux line and
$\nu_\pm$ is the dynamical phase. Henceforth, we will talk only
about the Berry phase and not the dynamical phase as it is
irrelevant in the context of the present study.
We may locate the Berry phase by decomposing the initial entangled singlet state into the eigenstates of the interaction Hamiltonian. Let us consider
the state
\begin{equation}
|\psi(t=0)>=a|\uparrow_n\downarrow_n>-b |\downarrow_n\uparrow_n>
\end{equation}
After one complete rotation the state picks up the geometric phase (apart from the dynamical phase which is omitted here)
\begin{equation}
|\psi(t=\tau)>=a e^{i\gamma_+}|\uparrow_n\downarrow_n>-be^{i\gamma_-} |\downarrow_n\uparrow_n>
\end{equation}
Geometrically, we may
get the explicit values of the Berry phase given by $\gamma_\pm$ as
\begin{equation}\label{gpn}
    \begin{array}{cl}
      \displaystyle{\gamma_+(\theta)}~= & \displaystyle{-\pi(1-\cos \theta)} \\
      &\\
      \displaystyle{\gamma_-(\theta)}~= & \displaystyle{-\pi(1+\cos \theta)=-\gamma_+(\theta)-2\pi} \\
    \end{array}
\end{equation}
Here the angle $\theta$ measures the deviation of the magnetic flux line from the z-axis.

As a fermion is depicted as a scalar particle encircling a magnetic
flux line which induces the Berry phase $e^{i2\pi \mu}$ with
$\mu=1/2$ we see from the above equations  that in an entangled
state the factor $\mu$  varies with the change of angle $\theta$. We
find that $\mu$ corresponds to a value
\begin{equation}\label{gpnm}
    \begin{array}{cl}
      \displaystyle{\gamma_+(\theta)~\rightarrow~\tilde{\mu}_+}~= &
      \displaystyle{-\frac{1}{2}(1-\cos \theta)} \\
      &\\
      \displaystyle{\gamma_-(\theta)~\rightarrow~\tilde{\mu}_-}~= &
      \displaystyle{-\frac{1}{2}(1+\cos \theta)} \\
    \end{array}
\end{equation}
Actually, the Berry phase factor $\mid \tilde{\mu}_{\pm}\mid
=\frac{1}{2}(1 \mp\cos \theta)$ gives the measure of formation of
entanglement.

It is noted that for $\mid \tilde{\mu}_{+}\mid~
~=\frac{1}{2}(1 -\cos \theta)$, $\theta=0$ and $\theta=\pi$ corresponds to the disentangled state and a state of maximal entanglement (MES)respectively. Indeed when $\theta=0$ (i.e. there is no displacement of the direction of the magnetic flux line from the z-axis) we have $\mid \tilde{\mu}_{+}\mid~=0$  which means disentanglement. For,$\theta=\pi$, i.e. there is maximum deviation of the magnetic flux line we have  $\mid \tilde{\mu}_{+}\mid~=1$ which suggests the maximum entanglement. Thus the value $\mid \tilde{\mu}_{+}\mid~$ within the range $0\leq \mid \tilde{\mu}_{+}\mid \leq 1$ gives the measure of entanglement.

We can find a relationship of concurrence with this factor $\mid \tilde{\mu}_{+}\mid$. The standard concurrence for the state (7) is given by \cite{wooters}
\begin{equation}\label{Cs}
    C=2|a|~|b|
\end{equation}
and for maximum entanglement we have $C=1$. 
In our formalism, the entanglement is considered to be caused by the displacement of the magnetic flux line associated with one particle under the influence of the other particle.
So, we have to consider $\mid a\mid $ and $\mid b\mid $ as functions of $\theta$. We can write,
\begin{equation}
\frac{1}{\sqrt 2}\left(
\begin{array}{c}
|a|\\
|b|
\end{array}
\right) =\left(
\begin{array}{c}
f(\theta)\\
g(\theta)
\end{array}
\right)
\end{equation}
We know that at the maximum entangled state (MES) the normalization condition is 
$$ \mid a\mid^2 ~+~\mid b\mid^2~=1 $$ and the concurrence is
$$ C=2\mid a\mid \mid b\mid~=1$$.
So we have,
\begin{equation}
\mid a\mid =\mid b\mid =\frac{1}{\sqrt 2}
\end{equation}  
This implies that,
\begin{equation}
f(\theta)\mid_{\theta=\pi}=~g(\theta)\mid_{\theta=\pi}=\frac{1}{2}
\end{equation}
Again, from the no entanglement condition $C=0$ at $\theta=0$ we have,
\begin{equation}
\rm{either}~f(\theta)\mid_{\theta=0}=0~~~~~~~~\rm{or}~g(\theta)\mid_{\theta=0}=0
\end{equation}
From these constraint equations, for the positive definite norms 
$0\leq \mid a \mid \leq 1$ and $0\leq \mid b \mid \leq 1$, we can write 

\begin{equation}
\frac{1}{\sqrt{2}}\left(
\begin{array}{c}
|a|\\
|b|
\end{array}
\right) =
\left(
\begin{array}{c}
f(\theta)\\
g(\theta)
\end{array}
\right)=
\left(
\begin{array}{c}
\cos^2\frac{\theta}{4}\\
\sin^2\frac{\theta}{4}
\end{array}
\right)
\end{equation}
The standard concurrence $C$ is found to be given by
 \begin{equation}
C=2|a|~|b|=\sin^2\frac{\theta}{2}=\frac{1}{2}(1-\cos\theta)=|\mu_{+}| \end{equation}
Thus we can formulate a relation between the concurrence $C$ and the geometric phase $e^{i\phi_B}$ in an entangled system. The relation is 
\begin{equation}\label{c1}
C=\frac{|\phi_B|}{2\pi}
\end{equation}
To substantiate our result let us now calculate the geometrical phase in two different spin systems and compare with the value of concurrence obtained in the literature.\\

{\bf{ Transverse Ising Model}}

The Hamiltonian for the anisotropic XY- model on 1D lattice with $N$ sites in a transverse field is given by
\begin{equation}
H=-\sum_{i=-M}^M \lambda\left[ \left(\frac{1+\gamma}{2}\right)\sigma_i^x\sigma_{i+1}^x
+~ \left( \frac{1-\gamma}{2}\right)\sigma_i^y\sigma_{i+1}^y+~\sigma_i^z \right]
\end{equation}
where $\sigma_i$ is the Pauli matrix at site $i$, $\gamma$ is the degree of anisotropy and $\lambda$ is the inverse strength of the external field. Here $M=\frac{N-1}{2}$ for $N$ odd.
For $\gamma=1$ the system corresponds to the transverse Ising model. Osborne et al \cite{osb} as well as Osterloh et al \cite{ost}have studied the concurrence of this system. They have also studied the critical behaviuor related to this. On the otherhand, the geometric phase of this model has been studied to understand the criticality in terms of this phase \cite{car,zhu}.
Following these works, we can calculate the $geometric$ $phase$ at the critical point and show its relation with the value of $concurrence$ at that point.

A family of Hamiltonians is introduced by applying a rotation of $\phi$ around the $z$-direction to each spin i.e. 
\begin{equation}
H_\phi=g_\phi~H~ g_\phi^\dagger
\end{equation} 
with $$g_\phi=\exp[\frac{i\phi \sigma_i^z}{2}]$$.
This class of models can be diagonalised by means of Jordan-Wigner transformation that maps spins to an one dimensional spinless fermions with creation and annihilation operators $a^\dagger_j$ and $a_j$ where
\begin{equation}
a_j=\left( \prod_{i<j}\sigma_i^z \right) \sigma_j^\dagger
\end{equation}
The Fourier transform of the fermionic operator is given by
\begin{equation}
d_k=\frac{1}{\sqrt{N}}\sum_j a_j \exp \left( -\frac{i 2 \pi j k}{N} \right)
\end{equation}
with $k=-M, -M+1....,M$. The Hamiltonian $H_\phi$ can be diagonalised by transforming the fermion operators in momentum space and then using the Bogoliubov transformation. From the procedure we can obtain the ground state $|g>$ given by 
\begin{equation}
|g>=-\prod_{k>0}(\cos\frac{\theta_k}{2}|0>_k|0>_{-k}-ie^{2i\phi}
\sin\frac{\theta_k}{2}|1>_k|1>_{-k})
\end{equation} 
where $|0>_k$ and $|1>_k$ are the vacuum and single fermionic excitation of the $k-th$ mode respectively. The angle $\theta_k$ is defined by
\begin{equation}
\cos \theta_k=\frac{1+\lambda \cos\phi_k}{\sqrt{1+\lambda^2+2\lambda\cos\phi_k}}
\end{equation}
with $\cos\phi_k=\frac{2\pi k}{N}$, $k=-M,-M+1,.....+M$.
\\
The geometric phase of the ground state accumulated by varying the angle $\phi$ from $0$ to $\pi$ is described by
\begin{equation}
\Gamma=-i \int_0^\pi~<g|\frac{\partial}{\partial \phi}|g> d\phi
=\sum_{k>0}\pi (1-\cos\theta_k)
\end{equation}
In the thermodynamic limit we have
\begin{equation}
\Gamma= \int_0^\pi~\left(1-\frac{1+\lambda \cos\phi}{\sqrt{1+\lambda^2+2\lambda\cos\phi}}
\right) d\phi
\end{equation}
which at the critical point $\lambda=\lambda_C=1$ gives,
\begin{equation}
\Gamma_C=\pi-2
\end{equation}
Referring to eqn.(\ref{c1}) we have the value of the  concurrence for the nearest neighbour
entanglement at the critical point
\begin{equation}
C=\frac{|\Gamma_C|}{2 \pi}=.18
\end{equation}
This is in very good agreement with the value of the nearest neighbour concurrence at the critical point obtained by the standard procedure $C=.1946$\cite{osb}

We should point out here that at other points of $\lambda$, the concurrence will depend on a function of $\lambda$ which is related to the external field. Indeed, as Carollo and Paschos \cite{car} have shown, the geometric phase is a witness of the singular point. 
Another point to be noted is that the relevance of the relation between concurrence and the geometrical phase becomes much more pronounced by the fact that near the critical point both concurrence and the Berry phase follow the scaling behaviour as has been shown in refs. \cite{ost,zhu} respectively.  

{\bf{Linear antiferromagnetic spin system}}

We consider the Hamiltonian for the Heisenberg antiferromagnetic spin model in a linear chain
\begin{equation}
H=\sum_{i,j}\frac{1}{2} (S_i^+~S_j^-)~+~ S_i^z S_j^z
\end{equation}
The sum is over all nearest neighbour pairs on a given lattice. The rotational symmtery of the Hamiltonian would imply for the $S=0$ sector ($S$=total spin)
\begin{equation}
<S_i^x S_j^x>=<S_i^y S_j^y>=<S_i^z S_j^z>
\end{equation}
For the spin-$1/2$ chain
\begin{equation}
|<S_i^z S_j^z>|\leq \frac{1}{4}
\end{equation}   
 Indeed we can write
 \begin{equation}
|<S_i^z S_j^z>|=\frac{1}{4}\cos\theta_{ij}
\end{equation}    
as under the influence of the other spin, the spin at a site deviates from its quantization axis
by an angle $\theta$.\\
 In the $S=0$ sector, we have the correlation
 \begin{equation}\label{gs}
 |<{\bf S}_i{\bf  S}_j>|=\frac{3}{4}\cos\theta_{ij}
\end{equation} 
 for the spin vectors ${\bf S}_i$ and ${\bf S}_j$.
 The Berry phase obtained by a rotation of the spin around the z-axis as shown in eqn.(9) is given by
 \begin{equation}\label{gl}
 \Gamma=\pi(1-\cos\theta)
 \end{equation}
  In a spin system, in the ground state the angle $\theta$ represents the deviation of one spin axis under the influence of another in the chain and so we can relate the Berry phase $\Gamma_{af}$ 
with the correlation $|<{\bf S}_i{\bf  S}_j>|$

In the linear antiferromagnetic chain, we find from eqns. (\ref{gs}) and (\ref{gl}), the Berry phase accumulated is given by
\begin{equation}
 \Gamma_{af}~=~\sum_{<i,j>}(\pi(1-4|<{\bf S}_i{\bf  S}_j>|)
 \end{equation}.  
 where $i$ and $j$ are nearest neighbour sites.
 The ground state energy of the Heisenberg Hamiltonian is given by
 \begin{equation}\label{eg}
 E_g=e_g  N=3N_{nn}\Gamma_g
\end{equation}
where $\Gamma_g$ is the nearest neighbour correlation $<S_i^z S_j^z>$, $e_g$ is the ground state energy per site and $N_{nn}$ is the number of nearest neighbour pairs of spins. 
  From(\ref{eg}), we have 
  \begin{equation}\label{eg1}
\Gamma_g= \frac{e_g . N}{3N_{nn}}
\end{equation} 
We have
\begin{equation}
3\cos\theta=4e_g\frac{N}{N_{nn}}
\end{equation}

The ground state energy per site $e_g= ln 2-\frac{1}{4} $ and for the thermodynamic limit
$N_{nn}=N$ we get 
\begin{equation}
\Gamma_{af}=\pi(1-4 e_g)
\end{equation} 
  Therefore,
   the concurrence for the pair formed by the nearest neighbour spins is given by
  \begin{equation}
  C=\frac{|\Gamma_{af}|}{2\pi}=\frac{1}{2}|(1-4e_g)|=2ln2-1=0.386
  \end{equation}
  This result is identical with that obtained by O'Connor and Wooters \cite{16}
  as well as Wang and Zanardi  \cite{17}. This also suggests the relationship between concurrence and antiferromagnetic correlations as envisaged by Subramanyam \cite{18}
  
  We can also note  from the relation(\ref{gl}) that for $\theta=\frac{\pi}{2}$ we have concurrence $C=0.5$ which happens in a frustrated spin system \cite{19}. This implies that the spin axis is orthogonal to the z-axis which induces the spin charge separation as is evident in high $T_c$ superconductors. Again when $\theta=0$ we have $C=0$ which corresponds to a ferromagnetic chain implying disentanglement.
  
  In summary we may say that from this geometrical point of view the concurrence of an entangled state is related to its geometrical phase. In this regard we can also suggest that the spin degrees of freedom of individual fermions can be entangled though the fermions themselves are noninteracting. This analysis implies that spin entanglement is associated with the spatial entanglement between electrons at different spatial points and as Vedral \cite{21}has pointed out, entanglement is a consequence of Fermi statistics.    
Indeed, in a separate note we have studied \cite{25}the entanglement of two identical fermions in two different spatial regions in terms of their spin entanglement and have found that the result is consistent with our present view point.

  \end{document}